\documentclass{optica-article}

\journal{opticajournal}
\articletype{Research Article}
\usepackage{multirow}
\usepackage{lineno}
\begin{document}

\title{Manipulation of polarization topology using a Fabry-P\'erot fiber cavity with a higher-order mode optical nanofiber}

\author{Maki Maeda,\authormark{1,*} Jameesh Keloth,\authormark{1} and S\'ile {Nic Chormaic}\authormark{1}}

\address{\authormark{1}Okinawa Institute of Science and Technology Graduate University, Onna, Okinawa 904-0495, Japan}

\email{\authormark{*}maki.maeda@oist.jp} %% email address is required; see note below about the corresponding author designation

% \homepage{http:...} %% author's URL, if desired

%%%%%%%%%%%%%%%%%%% abstract %%%%%%%%%%%%%%%%
%% [use \begin{abstract*}...\end{abstract*} if exempt from copyright]

\begin{abstract}
Optical nanofiber cavity research has mainly focused on the fundamental mode. Here, a Fabry-P\'erot fiber cavity with an optical nanofiber supporting the higher-order modes, TE$_{01}$, TM$_{01}$, HE$_{21}^o$, and HE$_{21}^e$, is demonstrated. Using cavity spectroscopy, with mode imaging and analysis, we observe cavity resonances that exhibit complex, inhomogeneous states of polarization with topological features containing Stokes singularities such as C-points, Poincar\'e vortices, and L-lines. \textit{In situ} tuning of the intracavity birefringence enables the desired profile and polarization of the cavity mode to be obtained. These findings open new research possibilities for cold atom manipulation and multimode cavity quantum electrodynamics using the evanescent fields of higher-order mode optical nanofibers.
\end{abstract}

%%%%%%%%%%%%%%%%%%%%%%%%%%  body  %%%%%%%%%%%%%%%%%%%%%%%%%%

\section{Introduction}

\label{sec:introduction}
Novel phenomena that can be revealed in non-paraxial light, such as transverse spin and spin-orbit coupling, have led to increasing interest in the tightly confined light observed in nano-optical devices \cite{Bliokh2015}. Optical nanofibers (ONFs), where the waist is subwavelength in size, are useful in this context because they provide very tight radial confinement of the electric field and facilitate diffraction-free propagation over several centimeters \cite{Solano2017}. Most ONF research focuses on single-mode ONFs (SM-ONFs) that only support the fundamental mode, HE$_{11}$. In contrast, higher-order mode ONFs (HOM-ONFs), fabricated from a few-mode optical fiber, can guide HOMs, such as TE$_{01}$, TM$_{01}$, HE$_{21}^e$, and HE$_{21}^o$ \cite{Frawley2012}. In the weakly guided regime, which is generally used to describe light propagation in standard optical fiber, this group of modes can be viewed to form the linearly polarized mode, LP$_{11}$. To date, there has been a lot more attention paid to HOM-ONFs in theoretical work \cite{Phelan2013, LeKien2017, LeKien2017b, LeKien2018a, LeKien2018b, Stourm2020, LeKien2022} than experimental work due to the difficulty in precisely controlling the fiber waist size and obtaining selective mode excitation at the waist \cite{Frawley2012, Hoffman2015, Fatemi2017}.

In principle, there are many interesting phenomena which can be explored with a HOM-ONF. For example, it has been proposed that the relationship between spin angular momentum (SAM) and orbital angular momentum (OAM) can be studied \cite{LeKien2017, Picardi2018, LeKien2019, LeKien2022}. Additionally, it was proposed that a HOM-ONF could be used to trap and manipulate cold atoms \cite{Sague2008, Phelan2013, Sadgrove2016}. Fabrication of an ONF that supports the HOMs was achieved \cite{Petcu-Colan2011, Frawley2012, Ward2014} and subsequently shown to more efficiently manipulate dielectric microbeads in the evanescent field than SM-ONFs \cite{Maimaiti2015, Maimaiti2016}. Other experimental work has shown that when cold atoms also interact with HOMs, detected signals are stronger than when one uses a SM-ONF only \cite{Kumar2015}. 

Introducing a cavity system to the ONF could further increase light-matter interactions due to cavity quantum electrodynamics (cQED) effects \cite{LeKien2009, Nayak2018, Romagnoli2020}. To date, numerous types of SM-ONF-based cavities have been proposed \cite{Keloth2017, Li2017, Li2018, Tashima2019, Ruddell2020, Li2021} and the interactions of their resonance modes with various quantum emitters have been studied \cite{Yalla2014, White2019, Tashima2022}. Strong light-atom coupling using SM-ONF-based Fabry-P\'erot and ring resonators has already been achieved \cite{Kato2015, Ruddell2017}. Superstrong coupling of cold atoms and multiple longitudinal modes of a long fiber-ring resonator consisting of a SM-ONF section was demonstrated \cite{Johnson2019}. Utilizing multiple degenerate higher-order transverse modes in free-space has shown to exhibit strong coupling \cite{Salzburger2002, Wickenbrock2013}, further illustrating the importance of realizing a HOM-ONF-based cavity system at this point. The advantages are not only for enhanced interactions via cQED effects, but also for a better overall understanding of the behavior of the modes in such a cavity. 

Studying the behavior of the HOM-ONF cavity spectrum and the cavity mode profiles gives additional insight into the nature of the HOMs themselves, as well as how they interfere with each other and interact with the external environment. The generation of TE$_{01}$ and TM$_{01}$ modes in a laser cavity consisting of a microfiber directional coupler-based mode converter was demonstrated previously \cite{Mao2018}. However, earlier attempts to realize a passive HOM optical microfiber cavity did not yield any resonant peaks in the cavity spectrum apart from the fundamental modes; in other words, the typical donut- or lobe-shaped intensity profiles associated with HOMs were not observed \cite{Jockel2009}, primarily due to challenges when engineering the taper profile to minimize losses at the taper transitions. 

The inhomogeneous polarization structure of HOMs needs to be taken into account when studying a fiber cavity system with a HOM-ONF. In recent years, complex polarization distributions and the generation of polarization singularities have been investigated using various methods, giving rise to the relatively new field of singular optics \cite{Wang2021}. Polarization singularities are a subset of Stokes singularities, \textit{i.e.}, phase singularity points in Stokes phases \cite{Freund2002, Arora2022}. In fact, higher-order fiber eigenmodes are vector optical fields with a polarization singularity called a V-point, where the state of polarization (SOP), \textit{i.e.}, how the polarization is distributed in the cross-section of a given mode, is undefined \cite{Wang2021}. Other types of Stokes singularities can be formed in elliptical optical fields, such as the polarization singularity of C-points, where the polarization orientation is undefined \cite{Freund2002, Wang2021}, and Poincar\'e vortices, where the polarization handedness is undefined \cite{Freund2001, Freund2002b, Arora2022}. Moreover, points of linear polarization can form continuous lines, which are classified as L-lines \cite{Wang2021}.

The generation of all Stokes singularities within a single beam has been demonstrated using a free-space interferometer \cite{Arora2019, Arora2022}. Modal interference in a birefringent crystal can facilitate the creation of polarization singularities \cite{Flossmann2005, Flossmann2006}. As a result, the SOP can significantly vary along the propagation length, with C-points and L-lines propagating as C-lines, \textit{i.e.}, continuous lines of circular polarization, and L-surfaces, \textit{i.e.}, surfaces of linear polarization, respectively \cite{Flossmann2005, Flossmann2006, Bliokh2008}. Moreover, polarization singularities can appear, move or disappear from a given cross-sectional region with a smooth and continuous change of birefringence \cite{Pal2017}. Birefringent media were used to create laser cavity modes containing a polarization singularity \cite{Pohl1971, Yonezawa2006}. These experiments were limited to the generation of low-order V-points due to a lack of control in the amplitude, phase, and SOP, all of which would be required to create other types of polarization singularities \cite{Wang2021}. A few-mode optical fiber cavity has the potential to generate complex laser modes by its highly variable degree of birefringence. 

\begin{figure}[htbp]
\centering
\includegraphics[width=\linewidth]{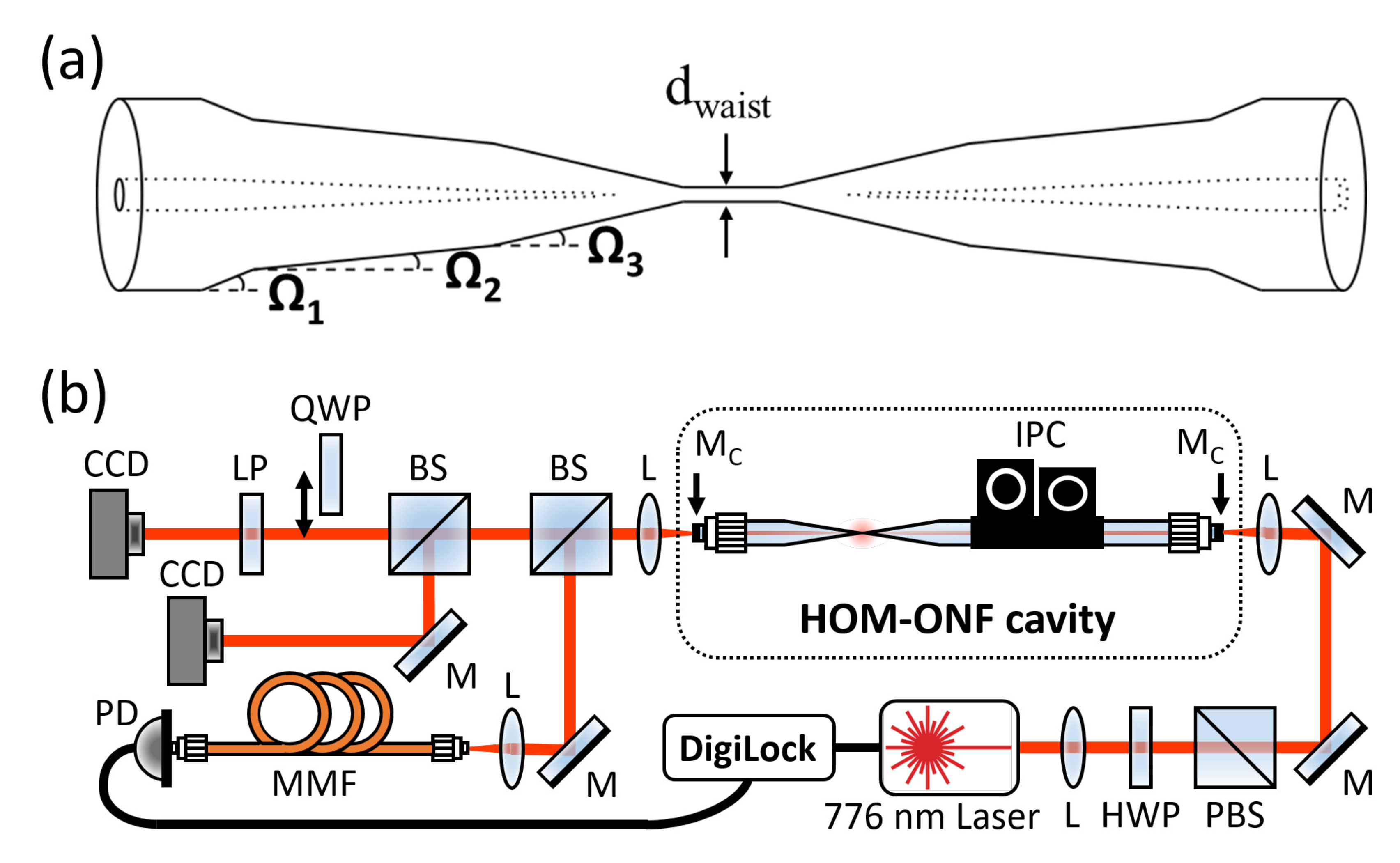}
\caption{(a) Sketch of tapered optical fiber with trilinear shape, d$_{waist}$: waist diameter. (b) Schematic of experimental setup. L: lens, HWP: half-wave plate, PBS: polarizing beam splitter, M: mirror, M$_{C}$: cavity mirror, IPC: in-line polarization controller, BS: beam splitter, QWP: quarter-wave plate, which was inserted to calculate S$_{3}$, LP: linear polarizer, CCD: camera, MMF: multimode fiber, PD: photodiode.}
\label{fig:ExpeimentalSetup}
\end{figure}

Interference and birefringence are generally inseparable properties in fibers. The modal interference pattern in a fiber changes continually with a periodicity of 2$\pi$ when the relative phase between modes is changed between 0 to 2$\pi$ as the eigenmodes propagate along the fiber \cite{Jiang2017}. This effect was used in a few-mode optical fiber to generate ellipse fields containing a C-point \cite{Jayasurya2011, Krishna2020}. Due to the increasing complexities of modal interference in few-mode fibers, filtering for the desired set of HOMs, and selectively exciting them to generate and manipulate polarization singularities, are necessary. Realizing a fiber cavity containing an ONF should enable both spatial and frequency filtering for selective excitation of HOMs, as well as enhancement of the resonant mode coupling effect \cite{LeKien2011, Keloth2019}.

In this paper, we experimentally demonstrate a HOM-ONF-based Fabry-P\'erot fiber cavity. The transverse polarization topology of any given resonant mode is determined by selecting modes from the cavity spectra and analyzing the images of the transmitted mode profile. We also demonstrate \textit{in situ} intracavity manipulation of the modal birefringence to change the amplitude, frequency position, and the SOP of the modes. This work is a significant step towards gaining full control of the evanescent field at the HOM-ONF waist and extends the range of applications for which such nanodevices could be used.

\section{Methods}
\label{sec:methods}

\subsection{Experiments}
\label{subsec:experiments}
For the HOMs described in Section \ref{sec:introduction} to propagate throughout the cavity with a HOM-ONF, the nanofiber must be low loss for the entire LP$_{11}$ set of modes. Tapered fibers were drawn from SM1250 (9/80) fiber (Fibercore) using an oxy-hydrogen flame pulling rig. The untapered fiber supports the LP$_{01}$, LP$_{11}$, LP$_{21}$, and LP$_{02}$ modes at a wavelength, $\lambda$~=~776~nm. The modes supported by the tapered fiber depend on the tapering profile and the waist diameter. We used two different tapered fibers with waist diameters of (i) $\sim$ 450~nm for SM behavior (HE$_{11}^o$ and HE$_{11}^e$) and (ii) $\sim$ 840~nm for the HOM-ONF, which supports HE$_{11}^o$, HE$_{11}^e$, TE$_{01}$, TM$_{01}$, HE$_{21}^o$, and HE$_{21}^e$. The shape of the tapered fibers was chosen to be trilinear, see Fig. \ref{fig:ExpeimentalSetup}(a), with angles of $\Omega_{1}$~=~2 mrad, $\Omega_{2}$~=~0.5~mrad and $\Omega_{3}$~=~1~mrad in order to be adiabatic for the LP$_{11}$ and LP$_{01}$ modes. Fiber transmission following the tapering process was >95$\%$ for the fundamental mode.

A sketch of the experimental setup is given in Fig. \ref{fig:ExpeimentalSetup}(b). The cavity was fabricated by splicing each pigtail of the tapered fiber to a commercial fiber Bragg grating (FBG) mirror (Omega Optical). The two FBG mirrors consisted of stacked dielectric mirrors coated on the end faces of fiber patchcords (SM1250 (9/80), Fibercore) and had a reflectivity of 97\% at $\lambda$~=~776~nm. Both mirrors had almost the same reflectivity over all input polarization angles (<~1\% variation). The cavity also contained an in-line polarization controller (IPC, see Fig.\ref{fig:ExpeimentalSetup}(b)) to manipulate the birefringence inside the cavity. Moving the paddles of the IPC induced stress and strain in the fiber, thereby changing the effective cavity length. A typical cavity length was $\sim$ 2~m, which was physically measured and estimated from the cavity free-spectral range (FSR).

A linearly polarized Gaussian beam from a laser at $\lambda$~=~776~nm (Toptica DL100 pro) was launched into the fiber cavity. The laser frequency was either scanned or locked to a mode of interest using a Pound-Drever-Hall locking module (Toptica Digilock110). The cavity output beam was split into three paths: one for the laser feedback controller to observe the cavity spectra and to lock to specific modes, one for imaging the spatial profile of the modes with a CCD camera, and one for analyzing the transverse SOP of each mode using a removable quarter wave plate (QWP), a rotating linear polarizer, and a CCD camera, see Fig. \ref{fig:ExpeimentalSetup}(b). Six intensity profile images were taken in total for each mode. Four images were taken without the QWP and with the linear polarizer angle set to 0$^{\circ}$ (I$_{H}$), 45$^{\circ}$ (I$_{D}$), 90$^{\circ}$ (I$_{V}$), and 135$^{\circ}$ (I$_{A}$), and two images were taken by inserting the QWP set to 90$^{\circ}$ while the polarizer was set to 45$^{\circ}$ (I$_{R}$) and 135$^{\circ}$ (I$_{L}$). The SOPs were determined by analyzing the six profile images using Stokes polarimetry. Furthermore, the Stokes phase and Stokes index were determined \cite{Wang2021}, see Section \ref{sec:methods} \ref{subsec:analysis}.

\subsection{Simulations}
\label{sec:theory}

Each mode experiences arbitrary birefringence as it propagates along the fiber. The total field in the fiber at any point is the sum of the propagating modes with a corresponding phase shift. The addition of FBG mirrors to the fiber induces an additional birefringence \cite{LeKien2011, Keloth2019}, which can be incorporated in a single birefringence matrix. Note, this model does not include cavity boundary conditions since we only aim to simulate the spatial profiles of the fiber modes. We can calculate an arbitrary fiber field, \textbf{E}, due to interference and birefringence by taking a summation over different fiber modes, such that
\begin{equation}
\textbf{E} = \sum_{M=1}^{n} J_{M}A_{M}\textbf{E}_{M}e^{i{\phi}_{M}},
\label{eq:interference+birefringence}
\end{equation}
where \textit{n} is the number of eigenmodes to be interfered, \textbf{E}$_{M}$ is the electric field of a fiber eigenmode \textit{M} $\in$ TE$_{0,m}$, TM$_{0,m}$, HE$_{\ell,m}$ and EH$_{\ell,m}$, with $\ell \in \mathbb{Z}^{+}$ being the azimuthal mode order, which defines the helical phase front and the associated phase gradient in the fiber transverse plane. $\textit{m} \in \mathbb{Z}^{+}$ is the radial mode order, which indicates the \textit{m}$^{th}$ solution of the corresponding eigenvalue equation \cite{LeKien2017}. \textit{A$_{M}$} is the amplitude, ${{\phi}_{M}}$ is the phase between modes, and \textit{J}$_{M}$ represents the arbitrary birefringence Jones matrix of each eigenmode \textbf{E}\textit{$_{M}$}, such that 

\begin{equation}
J_{M} = e^{i\eta_{M}/2}
\begin{pmatrix} 
cos^2\theta_{M} + e^{i\eta_{M}} sin^2\theta_{M} & (1-e^{i\eta_{M}}) cos\theta_{M} sin\theta_{M} \\
(1-e^{i\eta_{M}}) cos\theta_{M} sin\theta_{M}  & sin^2\theta_{M} + e^{i\eta_{M}} cos^2\theta_{M} \\
\end{pmatrix}
,
\label{eq:Jones matrix}
\end{equation}
where $\eta_{M}$ is the relative phase retardation induced between the fast axis and the slow axis, and $\theta_{M}$ is the orientation of the fast axis with respect to the horizontal-axis, \textit{i.e.}, perpendicular to mode propagation.

\begin{figure}[htbp]
\centering
\includegraphics[width=\linewidth]{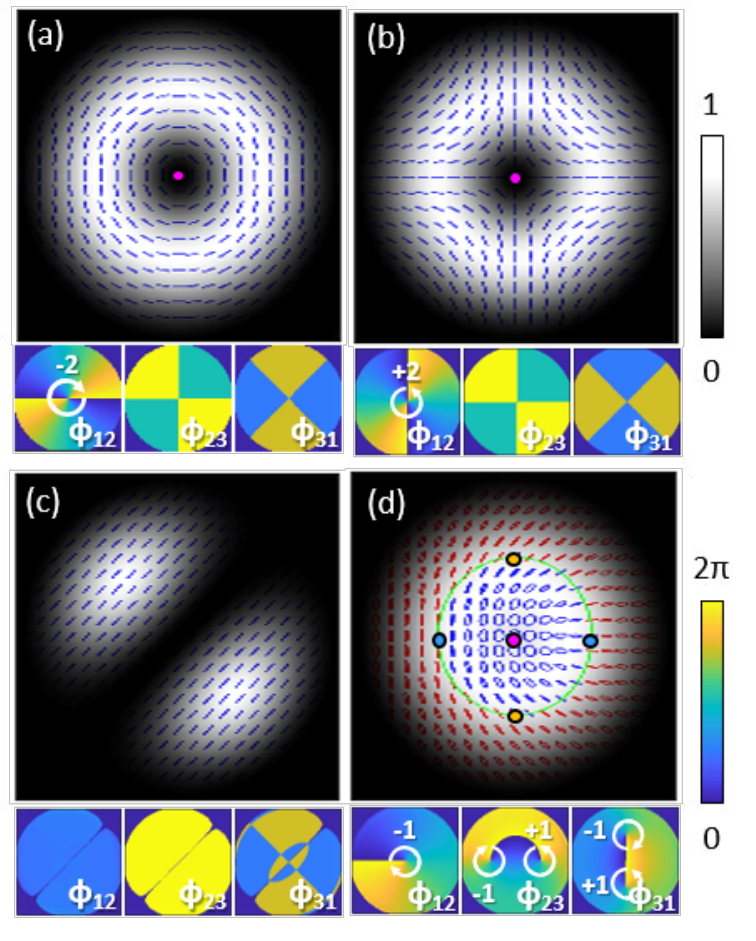}
\caption{Simulations of (a) TE$_{01}$, (b) HE$_{21}^{e}$, (c) TE$_{01}$ + HE$_{21}^{e}$ and (d) lemon. The red and blue SOPs indicate right-handed and left-handed ellipticities, respectively. The scale bars show the normalized intensity (from 0 to 1) and the Stokes phase (from 0 to 2$\pi$). Stokes singularity points of $\sigma_{12}$, $\sigma_{23}$, and $\sigma_{31}$ are indicated as pink, orange, and blue dots, respectively. An L-line is indicated in green.}
\label{fig:Simulations}
\end{figure}

Let us now consider the system with an ONF supporting HE$_{11}^o$, HE$_{11}^e$, TE$_{01}$, TM$_{01}$, HE$_{21}^o$ and HE$_{21}^e$, so that the number of modes that can be interfered is \textit{n}~$\leq$~6. The cross-sectional profiles and SOPs of TE$_{01}$ and HE$_{21}^e$ are shown in Fig. \ref{fig:Simulations}(a, b), respectively. The TM$_{01}$ and HE$_{21}^o$ modes are not shown here but their vector fields are orthogonal to the TE$_{01}$ and HE$_{21}^e$ at every point, respectively. These modes have donut-shape mode profiles with linearly polarized vector fields at any point in the mode cross-section. As an example of possible fiber modes using Eq. \ref{eq:interference+birefringence}, Fig. \ref{fig:Simulations}(c) illustrates in-phase interference of the TE$_{01}$ and HE$_{21}^e$ modes with equal amplitudes. The resulting mode has a lobe-shape intensity pattern with scalar fields. Fig. \ref{fig:Simulations}(d) is an example of a mode resulting from the interference of the circularly polarized HE$_{11}$ and an out-of-phase (a $\pi$/2 phase difference) TE$_{01}$ and TM$_{01}$ with equal amplitudes. The SOP, which is overlapped on the intensity profile images, are marked as red and blue ellipse, corresponding to right and left handed orientation, respectively. This mode is the co-called lemon \cite{Krishna2020}, which contains not only linear polarization but also elliptical and circular polarization components in one mode.

When using Eq. \ref{eq:interference+birefringence} to simulate mode profiles, a number of eigenmodes with similar intensity patterns and SOPs to an experimentally observed cavity mode were selected as the initial conditions. Next, the variables \textit{A}$_{M}$, $\phi_{M}$, $\eta_{M}$, and $\theta_{M}$ were tuned to match the experimentally observed cavity mode intensities, SOPs, and Stokes phases. Polarization topological defects in the simulated modes were then identified, using the method described in the following Section \ref{sec:methods} \ref{subsec:analysis}.

\subsection{Analysis}
\label{subsec:analysis}
The polarization gradient was calculated in order to identify Stokes singularities in the cross-section of the mode. The gradient map is known as the Stokes phase, $\phi_{ij}$, which is given by \cite{Freund2002, Freund2002b}
\begin{equation}
\phi_{ij} = {Arg}(S_{i} + iS_{j}),
\label{Stokes phase}
\end{equation}
where $S_{i}$ and $S_{j}$ are Stokes parameters with $\{\textit{i}, \textit{j}$\} $\in$ $\{1, 2, 3$\} in order, and \textit{i} $\neq$ \textit{j}. The phase uncertainty points, \textit{i.e.}, Stokes singularities, were identified by obtaining the Stokes indices, $\sigma_{ij}$, which are defined as \cite{Freund2002, Freund2002b}
\begin{equation}
\sigma_{ij} = \frac{1}{2\pi} \oint_{c} \phi_{ij} \cdot dc,
\end{equation}
where $\oint_{c}$ $\phi_{ij} \cdot dc$~=~$\Delta$ $\phi_{ij}$ is the counterclockwise azimuthal change of the Stokes phase around the Stokes singularity. Singularities of $\sigma_{12}$ are known as V-points and C-points, in vector and ellipse fields, respectively \cite{Freund2002}. Singularities of $\sigma_{23}$ and $\sigma_{31}$ are known as Poincar\'e vortices \cite{Freund2001, Freund2002b, Arora2022}. L-lines are located where $\phi_{23}$~=~$\{0, \pi, 2\pi\}$. Table \ref{tab:PS analysis} is a summary of the classification of the Stokes singularity types in terms of the Stokes phases and singularity indices with the corresponding polarizations in the vector and ellipse fields \cite{Freund2002b, Otte2018, Arora2019, Arora2022}.

\begin{table}[htbp]
\centering
\caption{\bf List of Stokes singularities in vector fields (v) and ellipse fields (e) by the singularity index, $\sigma_{ij}$, using the Stokes phase, $\phi_{ij}$, with \{\textit{i, j}\} $\in$ \{1, 2, 3\} in order.}
\begin{tabular}{cccc}
\hline
\multirow{2}{*}{Stokes} & \multirow{2}{*}{Stokes phase} & \multirow{2}{*}{Stokes index/} & \multirow{2}{*}{Polarization} \\
singularity &  & Phase values & \\
\hline\hline
V-point (v) & $\phi_{12}$ & $\sigma_{12}$ & Null\\
\hline
C-point (e) & $\phi_{12}$ & $\sigma_{12}$ & R/L\\
\hline
\multirow{2}{*}{Poincar\'e} & $\phi_{23}$ & $\sigma_{23}$ & H/V\\
\cline{2-4}
vortex (e) & $\phi_{31}$ & $\sigma_{31}$ & D/A\\
\hline
L-line (e) & $\phi_{23}$ & 0, $\pi$, 2$\pi$ & Linear\\

\hline
\end{tabular}
  \label{tab:PS analysis}
\end{table}

The Stokes singularity points and L-lines were found from the Stokes phases, then superimposed and marked on the mode profiles. As examples, from Figs. \ref{fig:Simulations}(a, b), the center of the mode profiles for both TE$_{01}$ and HE$_{21}^{e}$ contain a V-point, with $\sigma_{12}$~=~-2 and +2 (pink dot), respectively. These points were found from their Stokes phases $\phi_{12}$ (lower panels in Figs. \ref{fig:Simulations}(a, b)). In contrast, the lemon mode in Fig. \ref{fig:Simulations}(d) has a closed loop representing an L-line (green) and all three types of Stokes singularities: a C-point with $\sigma_{12}$~=~-1 (pink dot), Poincar\'e vortices with $\sigma_{23}$~=~-1 and +1 (orange dots), and $\sigma_{31}$~=~-1 and +1 (blue dots) were found from $\phi_{12}$, $\phi_{23}$, and $\phi_{31}$, respectively. The lobe-shaped scalar mode in Fig. \ref{fig:Simulations}(c) does not have a 2$\pi$ gradient in any associated Stoke phases, since topological defects can only exist in non-scalar fields \cite{Wang2021}.

\section{Results and discussion}
\label{sec:results}

\subsection{Cavity with a single-mode optical nanofiber}
\label{subsec:single-mode fiber cavity}
\begin{figure}[htbp]
\centering\includegraphics[width=\linewidth]{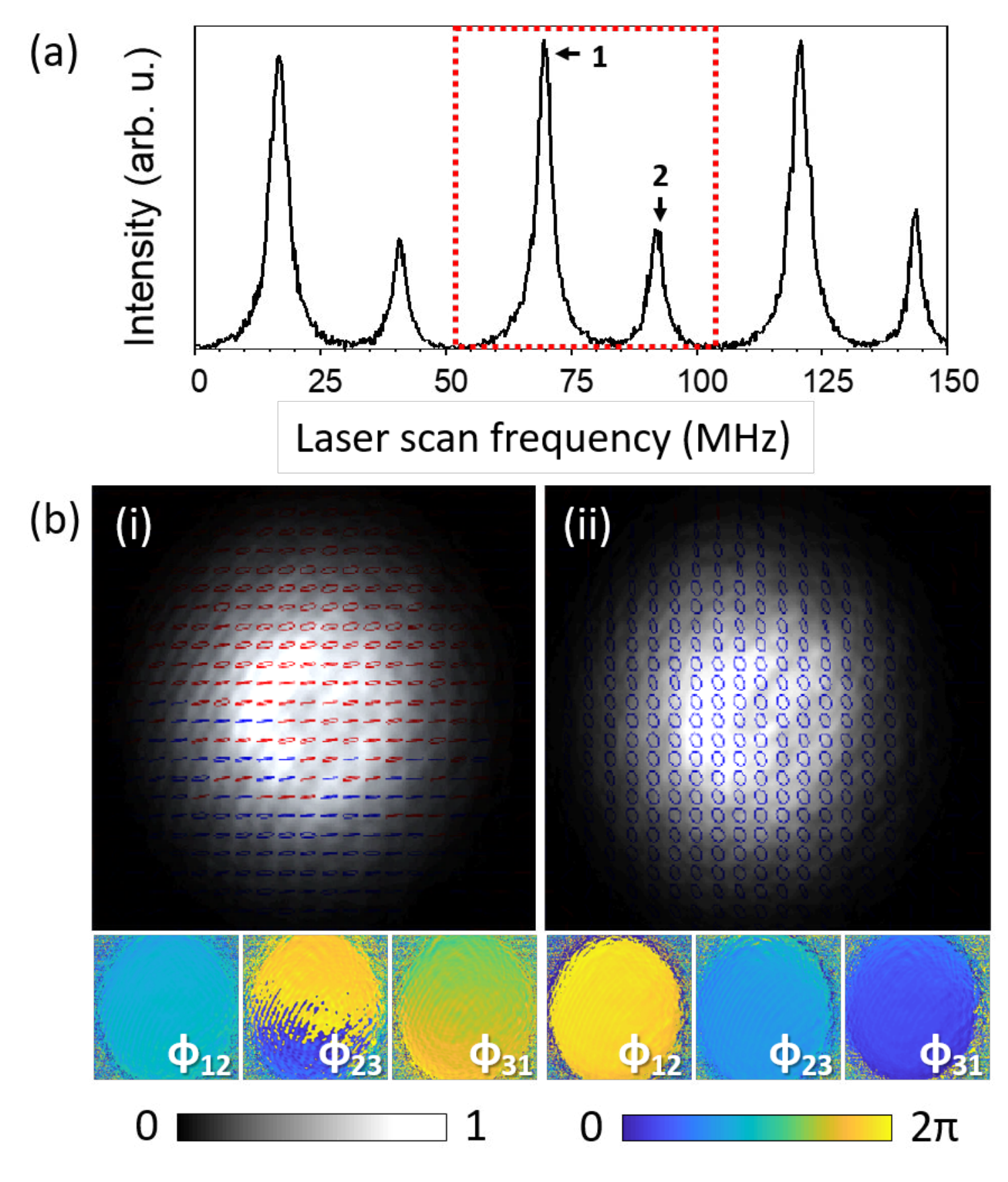}
\caption{(a) A typical spectrum for a HOM cavity with a SM-ONF as the laser is scanned over 150~MHz. The spectrum over a single FSR is indicated by the red box. (b) Mode intensity profiles showing the SOPs (top) and corresponding Stokes phases (bottom) for (i) Mode 1 and (ii) Mode 2. The red and blue SOPs indicate right-handed and left-handed ellipticities, respectively. The scale bars show the normalized intensity (from 0 to 1) and the Stokes phase (from 0 to 2$\pi$).}
\label{fig:SM_ONF}
\end{figure}

As an initial experimental test, the spectrum for a HOM cavity containing an ONF of waist diameter $\sim$ 450~nm was obtained, see Fig. \ref{fig:SM_ONF}(a). This ONF waist can only support the fundamental modes. The IPC paddle angles were set so that two distinct, well-separated modes with minimal spectral overlap were observed. The finesses of Modes 1 and 2 in Fig. \ref{fig:SM_ONF}(a) were 12 and 15, respectively. The laser was locked to each of these two cavity modes consecutively and the mode profiles were observed at the output end face of the fiber cavity. The corresponding mode intensity profiles, SOPs, and Stokes phases are shown in Figs. \ref{fig:SM_ONF}(b)(i, ii). The intensity profiles for both Modes 1 and 2 were slightly skewed Gaussian shapes.
The HE$_{11}$ eigenmode intensity shape is Gaussian, so the slight deviation from the expected shape may be attributed to aberrations in the optical beam path. In terms of polarization distribution, the Stokes phases of Modes 1 and 2 were uniform; in other words, their SOPs were scalar fields, regardless of the IPC paddle angles chosen, as expected for the HE$_{11}$ mode.

Although the pretapered fiber supported the full set of eigenmodes in LP$_{11}$, LP$_{02}$, and LP$_{21}$, when the ONF with a diameter $\sim$ 450~nm was inserted between the two sets of mirrors, only one or two modes with quasi-Gaussian profiles were observed, no matter which IPC paddle angles were chosen. The HOMs were filtered out due to the tapered fiber waist being SM, analogous to an intracavity pinhole spatial filter. Mode filtering as a function of the ONF waist diameter was observed experimentally \cite{Petcu-Colan2011}. However, here, we could additionally observe the mode filtering effect on the cavity spectrum and SOP of each mode.

\begin{figure*}[htbp]
\centering\includegraphics[width=\linewidth]{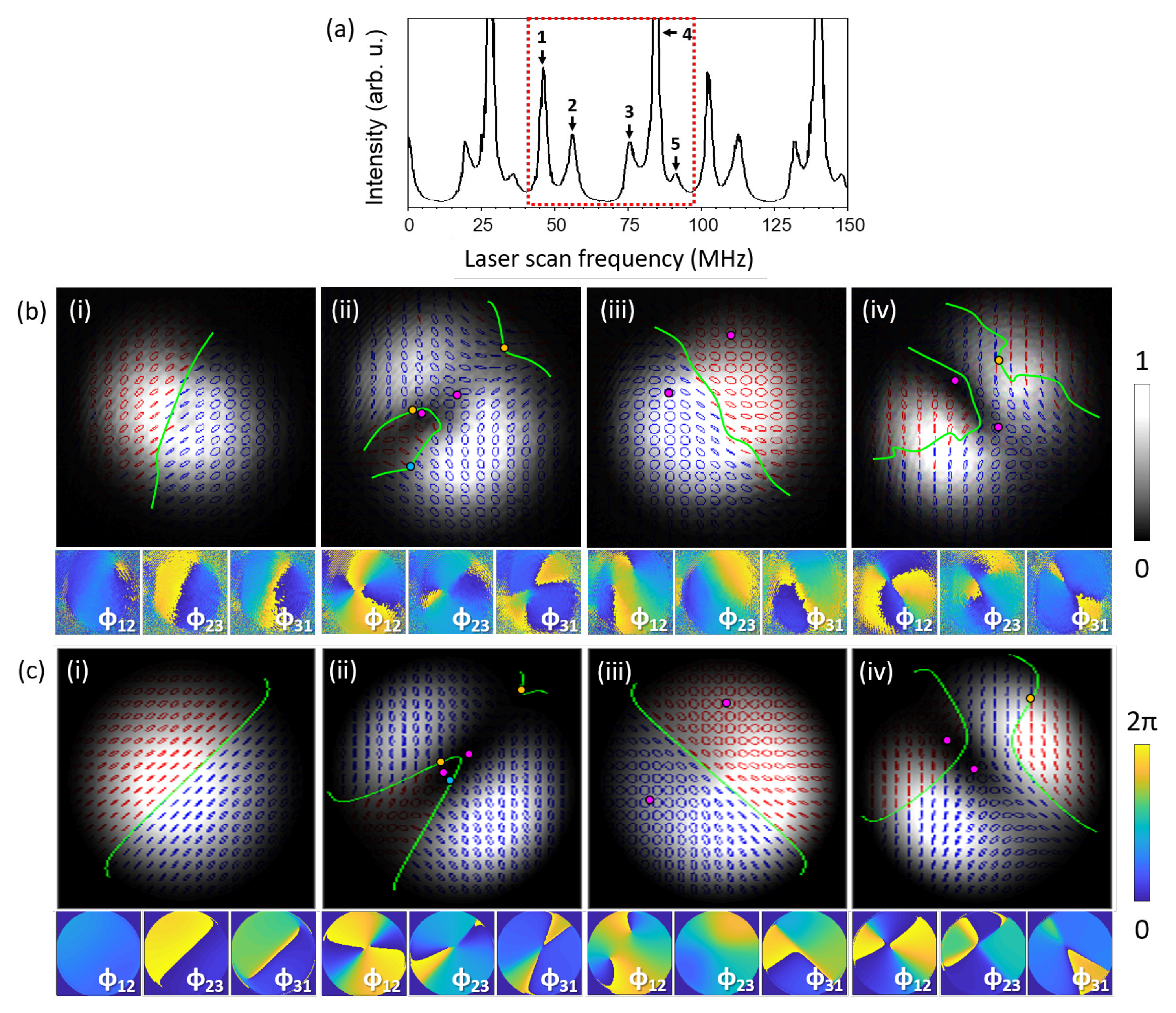}
\caption{(a) A typical spectrum for a cavity with a HOM-ONF as the laser is scanned over 150~MHz. The spectrum over a single FSR is indicated by the red box. (b) Mode intensity profiles showing the SOP (top) and the corresponding Stokes phases (bottom) for (i) Mode 1, (ii) Mode 2, (iii) Mode 4, and (iv) Mode 5. The red and blue SOPs indicate right-handed and left-handed ellipticities, respectively. The scale bars show the normalized intensity (from 0 to 1) and the Stokes phase (from 0 to 2$\pi$). Stokes singularity points of $\sigma_{12}$, $\sigma_{23}$, and $\sigma_{31}$ are indicated as pink, orange, and blue dots, respectively. L-lines are indicated in green. (c) Corresponding simulated results.}
\label{fig:HOM_ONF}
\end{figure*}

In an ideal SM-ONF cavity with no birefringence, there are two degenerate orthogonal modes. However, due to random birefringence of the fiber and the cavity mirrors, the two modes become non-degenerate, \textit{i.e.}, separated in frequency, leading to coupling between the modes \cite{Gordon2000}. Mode coupling of orthogonal modes can occur in a birefringent medium and this effect can increase in a cavity configuration \cite{Kolluru2018}. Mode coupling in an ONF cavity due to asymmetrical mirrors has been discussed previously \cite{LeKien2011} and experimental evidence of mode coupling due to intrinsic birefringence in a SM-ONF cavity has already been reported \cite{Keloth2019}. In our experiments, non-orthogonal combinations of SOPs were observed, as seen in Figs. \ref{fig:SM_ONF}(b)(i, ii). Mode 1 was horizontally polarized (red/blue lines in Fig. \ref{fig:SM_ONF}(b)(i)), while Mode 2 was left elliptically polarized (blue ellipse in Fig. \ref{fig:SM_ONF}(b)(ii)). By adjusting the IPC angles, it was possible to change the phase relationship and coupling between the HE$_{11}^o$ and HE$_{11}^e$ modes, and shift between orthogonal and non-orthogonal combinations of SOPs.

\subsection{Cavity with a higher-order mode optical nanofiber}
\label{subsec:few-mode fiber cavity}
Next, the spectrum for a HOM cavity containing an ONF of waist diameter $\sim$ 840~nm was obtained, see Fig. \ref{fig:HOM_ONF}(a). This ONF can support the HE$_{11}$, TE$_{01}$, TM$_{01}$, HE$_{21}^{o}$, and HE$_{21}^{e}$ modes. The IPC paddle angles were set to obtain the maximum number of well-resolved modes in a single FSR, see Fig. \ref{fig:HOM_ONF}(a). One can clearly see five distinct peaks indicating that the HOM-ONF does not degrade the modes in the cavity and the finesses of the cavity modes are high enough to resolve them. The finesses of Modes 1 to 5 were 12, 16, 13, 22, and 13, respectively. The mode finesse values of the cavity with a HOM-ONF were in the same range as those for the cavity with a SM-ONF (Fig. \ref{fig:SM_ONF}(a)), implying that the HOM-ONF was adiabatic for the LP$_{11}$ group of modes. The laser was locked to each of the cavity modes consecutively and the mode profiles were observed at the output of the fiber cavity. The corresponding mode intensity profiles, SOPs, and Stokes phases are shown in Figs. \ref{fig:HOM_ONF}(b)(i-iv). In the spectrum shown in Fig. \ref{fig:HOM_ONF}(a), there were five distinctive modes, but locking to Mode 3 was not possible because of its close proximity to the dominant Mode 4.

Two flat-top intensity profiles were observed in Modes 1 and 4, Figs. \ref{fig:HOM_ONF}(b)(i, iii) respectively. The SOPs of these modes are markedly different to those for the Gaussian-type modes in Figs. \ref{fig:SM_ONF}(b)(i, ii), which have simple scalar SOPs. Modes 1 and 4 were inhomogeneously polarized ellipse fields, showing regions of left and right circular polarizations divided by an L-line (Figs. \ref{fig:HOM_ONF}(b)(i, iii)). The center of these two modes exhibited diagonal and anti-diagonal polarizations, respectively, \textit{i.e.}, the SOPs at the center of the modes were orthogonal to each other. Going towards the edges of the modes, the polarization changes from linear to circular, with opposite handedness either side of the L-lines. Notice also in Fig. \ref{fig:HOM_ONF}(a) that Modes 1 and 4 are not well frequency separated from neighboring modes. This suggests that the mode profiles and SOPs of these modes were not only affected by birefringence and degenerate modal interference, but also some non-degenerate modal interference with neighboring cavity modes \cite{Kolluru2018}. Additionally, for Mode 4, we identified two C-points ($\sigma_{12}$~=~-1), indicated by the pink dots in Fig. \ref{fig:HOM_ONF}(b)(iii), where the value of $\phi_{12}$ changed by 2$\pi$ (see Table \ref{tab:PS analysis}). Interference of HE$_{11}$ with modes from the LP$_{11}$ group can generate C-points in a few-mode fiber \cite{Krishna2020}, see Fig. \ref{fig:Simulations}(d).

We performed basic simulations to determine if combinations of HE$_{11}$ and some mode(s) in the LP$_{11}$ family could generate similar mode profiles and SOP structures as those in Figs. \ref{fig:HOM_ONF}(b)(i, iii). The simulated results are shown in Figs. \ref{fig:HOM_ONF}(c)(i, iii). The HE$_{11}$ and TM$_{01}$ modes were selected as possible contributors and their amplitudes, phase, and birefringence fitting parameters were tuned to match the experimental results. Modes 1 and 4, see Figs. \ref{fig:HOM_ONF}(b)(i, iii), could have been formed from different mode combinations rather than our assumed HE$_{11}$ and TM$_{01}$; however, these modes were very likely formed by interference between HE$_{11}$ and some mode(s) of the LP$_{11}$ group, resulting in their inhomogeneous SOPs and flat-top shapes.

We also observed two distorted lobe-shaped modes, Modes 2 and 5, see Figs. \ref{fig:HOM_ONF}(b)(ii, iv). The lobe-shaped pattern also arises from modal interference between modes in the LP$_{11}$ family (as an example, see Fig. \ref{fig:Simulations}(c)). With reference to Table \ref{tab:PS analysis}, Mode 2, Fig. \ref{fig:HOM_ONF}(b)(ii), showed all three types of Stokes singularities, indicated by pink dots for C-points ($\sigma_{12}$~=~+1) and orange/blue dots for Poincar\'e vortices ($\sigma_{23}$~=~-1 /$\sigma_{31}$~=~+1), as presented in $\phi_{12}$, $\phi_{23}$, and $\phi_{31}$, respectively. A single mode containing all Stokes singularities has been demonstrated using free-space interferometers \cite{Arora2019, Arora2022}; here, we generated them within a single mode using a fiber cavity system. Mode 5, Fig. \ref{fig:HOM_ONF}(b)(iv), also had two C-points ($\sigma_{12}$~=~+1) and a Poincar\'e vortex ($\sigma_{23}$~=~+1), as seen in $\phi_{12}$, and $\phi_{23}$, respectively. Fig. \ref{fig:HOM_ONF}(a) shows that Modes 2 and 5 are not well frequency separated from Modes 1 and 4, respectively. Therefore, there is a likely contribution from the HE$_{11}$ mode resulting in distortion of the lobe shape.

To simulate Mode 2 in Fig. \ref{fig:HOM_ONF}(b)(ii), we combined TE$_{01}$, HE$_{21}^{e}$, and HE$_{11}$, and to simulate Mode 5 in Fig. \ref{fig:HOM_ONF}(b)(iv), we used TM$_{01}$, HE$_{21}^{e}$, and HE$_{11}$. The amplitude of each mode, phase shift, and birefringence parameters were adjusted to achieve a close fit. The simulated results are shown in Figs. \ref{fig:HOM_ONF}(c)(ii, iv).  These plots are not exact replications of the experimental results since the parameter space is large and the exact initial conditions are not known; nevertheless, the match is reasonably close.

Interestingly, many of the cavity modes obtained in different sets of spectra, which were generated using different IPC angles, exhibited Stokes singularities. Polarization singularities are known to propagate through a birefringent medium as C-lines and L-surfaces and their evolution is affected by the homogeneity of the birefringence along the propagation path \cite{Flossmann2005, Flossmann2006, Bliokh2008}. This phenomenon is due to the conservation of the topological charge \cite{Bliokh2008, Vyas2013, Otte2018}, and the Stokes index value, $\sigma_{ij}$, remains constant \cite{Otte2018}. However, our cavity is an inhomogeneous birefringent medium as it contains a number of different birefringent elements such as the FBG mirrors and the IPC, as such, the degree of birefringence varies along the propagation direction. Therefore, the presence of Stokes singularities in the imaged field at the cavity output does not necessarily guarantee the existence of such topological defects in the ONF region. Nonetheless, singularity points can enter, move and exit with a smooth and continuous variation of birefringence \cite{Pal2017}. Therefore, the SOP is expected to evolve along the length of the cavity, with singularity points shifting and making numerous entries and exits in the cross-section profile of the modes. However, since the ONF waist is relatively straight and uniform, the birefringence variation at the waist should be minimal \cite{Lei2019} and topological features appearing at the start of the waist should be preserved every 2$\pi$ along the waist.

Theoretically, the HOM-ONF can support a total of six eigenmodes as mentioned earlier. Therefore, one might expect that the spectrum should show six distinct modes. However, we typically observed three to five distinct peaks in a single FSR depending on the IPC paddle angles. This could be explained by the lack of sufficient finesse to resolve all modes, some of which are closely overlapped \cite{Kolluru2018}. However, it may be feasible to increase the mode finesses by increasing the mirror reflectivity and using an ONF with lower transmission loss than the one used (the estimated loss of Mode 4, the highest finesse in Fig. \ref{fig:HOM_ONF}(a), was $\sim$ 20$\%$). Nonetheless, the finesse values of our $\sim$ 2~m long cavity with a HOM-ONF should be sufficient for cQED experiments with narrow line-width emitters such as cold atoms.

\begin{figure*}[htbp]
\centering\includegraphics[width=\linewidth]{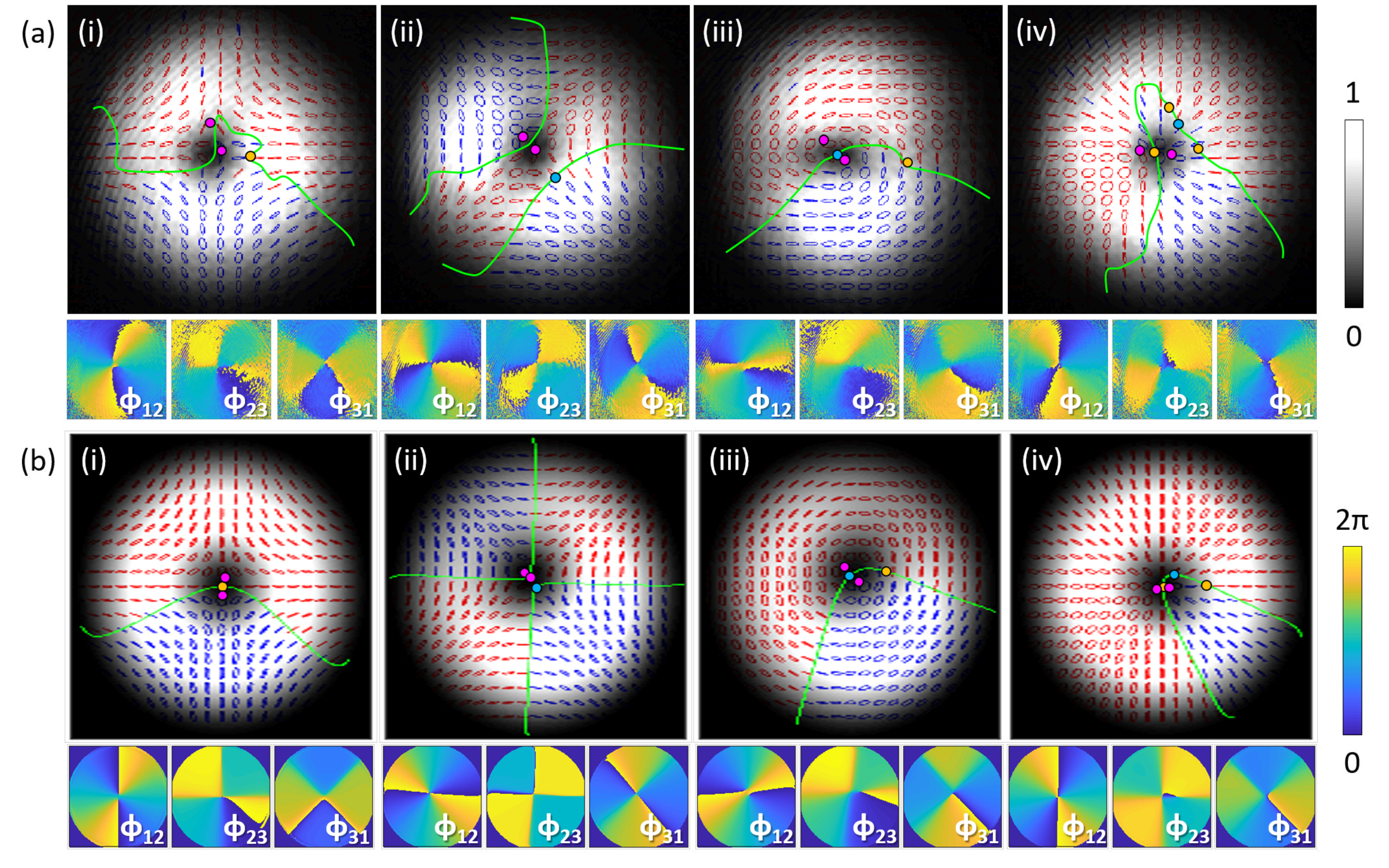}
\caption{(a) Mode intensity profiles for quasi-donut-shaped cavity modes from the cavity containing a HOM-ONF with their SOPs (top) and Stokes phases (bottom) similar to the fiber eigenmodes of (i) HE$_{21}^{e}$, (ii) HE$_{21}^{o}$, (iii) TE$_{01}$, and (iv) TM$_{01}$. The red and blue SOPs indicate right-handed and left-handed ellipticities, respectively. Scale bars show intensity (from 0 to 1) and Stokes phase (from 0 to 2$\pi$). Stokes singularities of $\sigma_{12}$, $\sigma_{23}$, and $\sigma_{31}$ are indicated as pink, orange, and blue dots, respectively. L-lines are illustrated as green lines. (b) Corresponding simulated results.}
\label{fig:quasieigenmode}
\end{figure*}

\subsection{\textit{In situ} higher-order cavity mode tuning}
\label{subsec:insitu}
A key feature of this setup is the ability to tune the spectrum and SOP to create the desired mode in the cavity. We aimed to observe modes with donut-shaped intensity patterns and SOPs similar to the fiber eigenmodes TE$_{01}$ (Fig. \ref{fig:Simulations}(a)), TM$_{01}$, HE$_{21}^{o}$, and HE$_{21}^{e}$ (Fig. \ref{fig:Simulations}(b)). To achieve this, the laser was locked to a well-resolved lobe-shaped mode. The paddle angles of the IPC were then adjusted, and the mode shape was monitored with a CCD camera until a donut mode profile was observed. Unlocking and scanning the laser revealed a new spectrum with each mode containing a new profile. The IPC was adjusted again to maximize another mode and the laser was locked to this new mode. The IPC paddle angles were tuned to once more convert the mode profile to a donut shape. This procedure was repeated for four different modes, see Figs. \ref{fig:quasieigenmode}(a)(i-iv), and these modes look similar to the true fiber eigenmodes of HE$_{11}^{e}$ (Fig. \ref{fig:Simulations}(b)), HE$_{11}^{o}$, TE$_{01}$ (Fig. \ref{fig:Simulations}(a)), and TM$_{01}$, respectively. There was a slight deformation from a perfect donut shape and their SOPs were not vector fields, but rather ellipse fields with alternating regions of opposite handiness. While the donut eigenmodes possessed a V-point at the center as indicated by pink dots in Figs. \ref{fig:Simulations}(a, b), the observed quasi-donut modes in Figs. \ref{fig:quasieigenmode}(a)(i-iv) had some nominal intensity at the center. These modes had two C-points of $\sigma_{12}$~=~-1 or +1 near the center (see pink dots in Figs. \ref{fig:quasieigenmode} (a)(i-iv)), as opposed to a single point of $\sigma_{12}$~=~-2 or +2 in the true eigenmodes (Figs. \ref{fig:Simulations}(a, b)). Indeed, perturbation of vector field polarization singularities can occur when scalar linearly polarized beams are interfered \cite{Arora2023}.

These donut-shaped cavity modes were also simulated, as shown in Figs. \ref{fig:quasieigenmode}(b)(i-iv). To obtain a good fit for the experimentally observed intensities, SOPs, and Stokes phases in Figs. \ref{fig:quasieigenmode}(a)(i-iv), the simulated modes included a slight deformation of the donut shape by adding some components of the HE$_{11}$ mode to modes in the LP$_{11}$ group. Moreover, the simulated results show that the Stokes phases are very similar to those obtained experimentally. The number of possible combinations of modal interference with varying birefringence is large and this leads to discrepancies between the experiment and simulation. However, these findings indicate that the experimentally observed quasi-donut modes are likely the result of residual interference between the HE$_{11}$ mode and modes in the LP$_{11}$ group. Degeneracy of multiple modes may be avoided by increasing the cavity mode finesses so that each mode can be well separated. The system demonstrated here shows that, even in a complex system, the HOMs and their SOPs can be controlled to create exotic topological states.

\section{Conclusion}
We have experimentally demonstrated a Fabry-P\'erot fiber cavity with a HOM-ONF and performed cavity spectroscopy. The cavity mode profiles and transverse polarization topology were also determined by imaging and analyzing the individual cavity modes at the output. These modes had inhomogeneous polarization distributions with a number of Stokes singularities. We also simulated the fiber modes which closely match those observed at the output of the cavity. Moreover, \textit{in situ} intracavity manipulation of the modal birefringence and interference to select a specific mode of interest was demonstrated. This indicates that the evanescent field of an HON-ONF could be tuned by adjusting the IPC paddle angles.

These findings are a step toward investigating the interactions between SAM and OAM of a HOM-ONF. Research into the interference of HOMs at the waist of an ONF is an exciting opportunity to uncover the nature of light-matter interactions in tightly confining geometries with topological singularities. Additionally, the realization of a (de)multiplexing system using degenerate HOMs in an ONF-based cavity may be possible by improving the tunability of the modal birefringence and interference. Such a system is attractive for future quantum information platforms as efficient and secure storage.

The interference of higher-order cavity modes with fixed ratios in the evanescent field of an ONF may also be used to trap and manipulate cold atoms. Adjusting the overlap and SOP of the HOMs should result in movement of the trapping sites relative to each other, enabling some trap dynamics to be studied \cite{Sague2008, Phelan2013, Sadgrove2016}. This cavity could be also used with quantum emitters to study multimode cQED effects using degenerate HOMs. The HOM cavity studied here had moderate finesse to enter the cQED experiments for interactions with cold atoms. In free-space optics, strong coupling of multiple transverse HOMs with atoms has been achieved \cite{Wickenbrock2013}, whereas this has not been achieved using an ONF-type cavity. Our work is a significant step towards this realization.

Moreover, the ability of our cavity to generate all three types of Stokes singularities may be useful to realize not only a C-point laser but also an all-Stokes singularity laser using a few-mode fiber. The combinations of fiber modes that we used in the simulations were found via manual trial-and-error estimates to obtain a visual match with the experimentally observed modes. More accurate control could be achieved by using machine learning techniques to fully cover the parameter space of permitted modes in the cavity. This may enable us to determine the correct combination of modes that lead to the observed cavity outputs and facilitate feedback to optimize the input to the system to generate desired modes in the cavity.

%\begin{backmatter}
%\bmsection{Funding} Okinawa Institute of Science and Technology Graduate University.

%\bmsection{Acknowledgments} The authors acknowledge F. Le Kien, L. Ruks, V. G. Truong, and J. M. Ward for discussions and K. Karlsson for technical assistance.

%\bmsection{Disclosures} The authors declare no conflicts of interest.

%\bmsection{Data availability} Data underlying the results presented in this paper are not publicly available at this time but may be obtained from the authors upon reasonable request.

%\end{backmatter}

%% Bibliography
%\bibliography{reference}

%\end{document}

\begin{backmatter}
\bmsection{Funding}
Okinawa Institute of Science and Technology Graduate University.

\bmsection{Acknowledgments}
The authors acknowledge F. Le Kien, L. Ruks, V. G. Truong, and J. M. Ward for discussions and K. Karlsson for technical assistance.

\bmsection{Disclosures}
The authors declare no conflicts of interest.

\bmsection{Data availability} Data underlying the results presented in this paper are not publicly available at this time but may be obtained from the authors upon reasonable request.

\end{backmatter}

%%%%%%%%%%%%%%%%%%%%%%% References %%%%%%%%%%%%%%%%%%%%%%%%%

%\bibliography{reference}

\end{document}